\documentclass[letterpaper,aps,prb,groupedaddress,preprint]{revtex4}

\usepackage{graphicx}

\begin{document}

\title{Entropy change at the martensitic transformation in ferromagnetic
shape memory alloys
$\mathbf{Ni}_{2+x}\mathbf{Mn}_{1-x}\mathbf{Ga}$}

\author{V.~V.~Khovailo}
\email[Electronic address: ]{vv-khovailo@aist.go.jp}
\author{K.~Oikawa}
\author{T.~Abe}

\affiliation{National Institute of Advanced Industrial Science and
Technology, Tohoku Center, Sendai 983--8551, Japan}

\author{T.~Takagi}
\affiliation{Institute of Fluid Science, Tohoku University, Sendai
980--8577, Japan}

\begin{abstract}
The entropy change $\Delta S$ between the high-temperature cubic
phase and the low-temperature tetragonally-based martensitic phase
of Ni$_{2+x}$Mn$_{1-x}$Ga ($x = 0 - 0.20$) alloys was studied. The
experimental results obtained indicate that $\Delta S$ in the
Ni$_{2+x}$Mn$_{1-x}$Ga alloys increases with the Ni excess $x$.
The increase of $\Delta S$ is presumably accounted for by an
increase of magnetic contribution to the entropy change. It is
suggested that the change in modulation of the martensitic phase
of Ni$_{2+x}$Mn$_{1-x}$Ga results in discontinuity of the
composition dependence of $\Delta S$.
\end{abstract}



\maketitle

\section{Introduction}

For shape memory alloys, the change of entropy $\Delta S$ between
high-temperature austenitic and low-temperature martensitic phase
can be obtained either from
calorimetry~\cite{1-planes,2-obrado,3-pelegrina} or from results
of stress-strain measurements at different temperatures above the
martensitic start temperature $M_s$ (Ref.~4). Owing to the
diffusionless character of martensitic transformations,
configuration contributions to the entropy change are absent,
which considerably simplify the evaluation of the relative phase
stability. In the case of thermoelastic martensitic
transformations, which are characterized by a small temperature
hysteresis and complete transformation to the austenitic
(martensitic) state, the change of entropy $\Delta S$ can be
determined experimentally with a good precision.

Ni$_2$MnGa, a representative of the family of Heusler alloys,
undergoes a thermoelastic martensitic transformation on cooling
below $T_m \sim 200$~K. Since ferromagnetic ordering in this
compound sets at a considerably higher temperature, $T_C = 376$~K,
the martensitic transformation occurs in the ferromagnetic state.
Both $T_m$ and $T_C$ are sensitive to stoichiometry. For instance,
a partial substitution of Mn for Ni in Ni$_{2+x}$Mn$_{1-x}$Ga
alloys results in increase of $T_m$ and decrease of $T_C$ until
they couple in a composition range $x = 0.18 - 0.20$ (Ref.~5).
Results of x-ray and electron diffraction studies of Ni-Mn-Ga
alloys indicate that the crystal structure of the martensitic
phase depends on composition. The martensitic phase of the alloys
with a low temperature of martensitic transformation ($T_m <
270$~K) has a five-layered modulation whereas the martensitic
phase with a moderate temperature of martensitic transformation
($T_m > 270$~K) has a seven-layered modulation.~\cite{6-pons} For
Cu-based shape memory alloys, which transform to various
martensitic structures upon cooling, it has been shown that the
entropy change depends on the particular structure of the
low-temperature martensitic phase.~\cite{2-obrado} Hence, similar
behavior could be expected in the Ni-Mn-Ga alloys. Contrary to the
Cu-based shape memory alloys, which are nonmagnetic, Ni-Mn-Ga
alloys possess a long-range ferromagnetic ordering at temperatures
below $T_C$. Such distinct magnetic properties could result in
peculiar behavior of the entropy change in Ni-Mn-Ga as compared to
nonmagnetic shape memory alloys.  The purpose of this work is to
perform a preliminary calorimetric analysis of the entropy change
$\Delta S$ between the high-temperature cubic phase and
low-temperature tetragonally based martensitic phases of
Ni$_{2+x}$Mn$_{1-x}$Ga ($x = 0 - 0.20$) alloys.

\section{Experimental details}

Polycrystalline ingots of Ni$_{2+x}$Mn$_{1-x}$Ga ($x = 0 - 0.20$)
alloys were prepared by an arc-melting method. The ingots were
annealed in evacuated quartz ampoules at 1050~K for 9~days. Sample
for calorimetric measurements were spark cut from the middle part
of the ingots. The calorimetric measurements were performed using
a Perkin-Elmer differential scanning calorimeter with a
heating/cooling rate of 5~ K/min. In the experiments we have also
used samples with the same thermal treatment from our previous
work.~\cite{7-kvv}

\section{Experimental results and discussion}

An example of the calorimetric measurements of
Ni$_{2+x}$Mn$_{1-x}$Ga alloys is presented in Fig.~1. The direct
and reverse martensitic transformations are accompanied by
well-defined calorimetric peaks. From these data, it is easy to
determine characteristic temperatures of the direct (martensite
start, $M_s$ and martensite finish, $M_f$) and the reverse
(austenite start, $A_s$ and austenite finish, $A_f$) martensitic
transformation. Results for the alloys studied, together with
composition of the samples and the equilibrium temperature $T_0 =
(M_s + A_f)/2$, are given in Table~I. It is worth noting that
transformation temperatures slightly differ for different samples
of the same composition and the values of the temperatures
presented in Table~I are averaged over several specimens.

\begin{figure}[t]
\begin{center}
\includegraphics[width=\columnwidth]{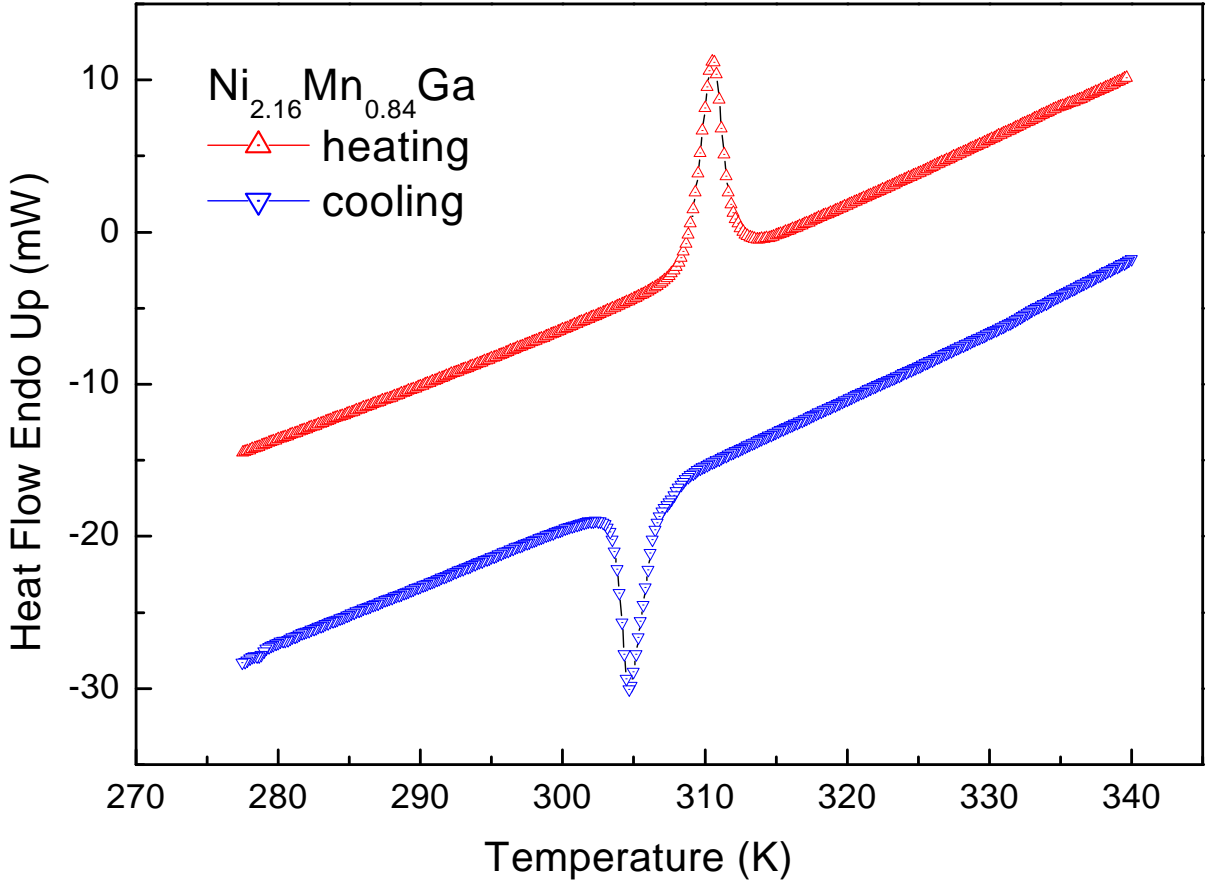}
\caption{Typical calorimetric curves corresponding to the direct
(cooling) and reverse (heating) martensitic transformations,
measured in Ni$_{2+x}$Mn$_{1-x}$Ga alloys.}
\end{center}
\end{figure}

\begin{table}[h]
\caption{Composition of the studied Ni$_{2+x}$Mn$_{1-x}$Ga alloys
and the critical temperatures of the martensitic transformation.}
\begin{ruledtabular}
\begin{tabular}{cccccc}

Alloy &$M_s$ (K) & $M_f$ (K) & $A_s$ (K) & $A_f$ (K) & $T_0$ (K)\\

\hline

\hline

$x = 0$ & 194 & 187 & 198 & 203 & 199 \\

\hline

$x$ = 0.02 & 221 & 214 & 224 & 229 & 225 \\

\hline

$x$ = 0.03 & 229 & 224 & 233 & 237 & 233 \\

\hline

$x$ = 0.04 & 238 & 233 & 238 & 243 & 240 \\

\hline

$x$ = 0.05 & 242 & 237 & 244 & 248 & 245 \\

\hline

$x$ = 0.08 & 266 & 262 & 269 & 272 & 269 \\

\hline

$x$ = 0.10 & 274 & 269 & 277 & 281 & 277 \\

\hline

$x$ = 0.13 & 277 & 272 & 280 & 285 & 281 \\

\hline

$x$ = 0.16 & 308 & 304 & 308 & 312 & 310 \\

\hline

$x$ = 0.18 & 329 & 324 & 332 & 337 & 333 \\

\hline

$x$ = 0.19 & 338 & 331 & 342 & 348 & 343 \\

\hline

$x$ = 0.20 & 338 & 332 & 344 & 349 & 344 \\

\end{tabular}
\end{ruledtabular}
\end{table}

The mean values of the heat exchanged upon the reverse ($Q^{L \to
H}$) and direct ($Q^{H\to L}$) transformation are shown in
Table~II. The average of the absolute values of ($Q^{L\to H}$) and
($Q^{H\to L}$) was taken as the change of enthalpy $\Delta H$.
When the Gibbs free energies of martensite and austenite are
equal, which takes place at temperature $T_0$, the entropy change
$\Delta S$ can be evaluated as $\Delta S = \Delta H/T_0$.
Determined in such a way, the entropy change is also shown in
Table~II.

Figure~2 shows the entropy change $\Delta S$ as a function of Ni
excess $x$ in the Ni$_{2+x}$Mn$_{1-x}$Ga alloys. It is evident
that $\Delta S$ increases with deviation from the stoichiometry.
It can also be inferred that the entropy change has different
composition dependencies in concentration intervals $0 \le x \le
0.13$ and $0.16 \le x \le 0.20$.

\begin{figure}[t]
\begin{center}
\includegraphics[width=\columnwidth]{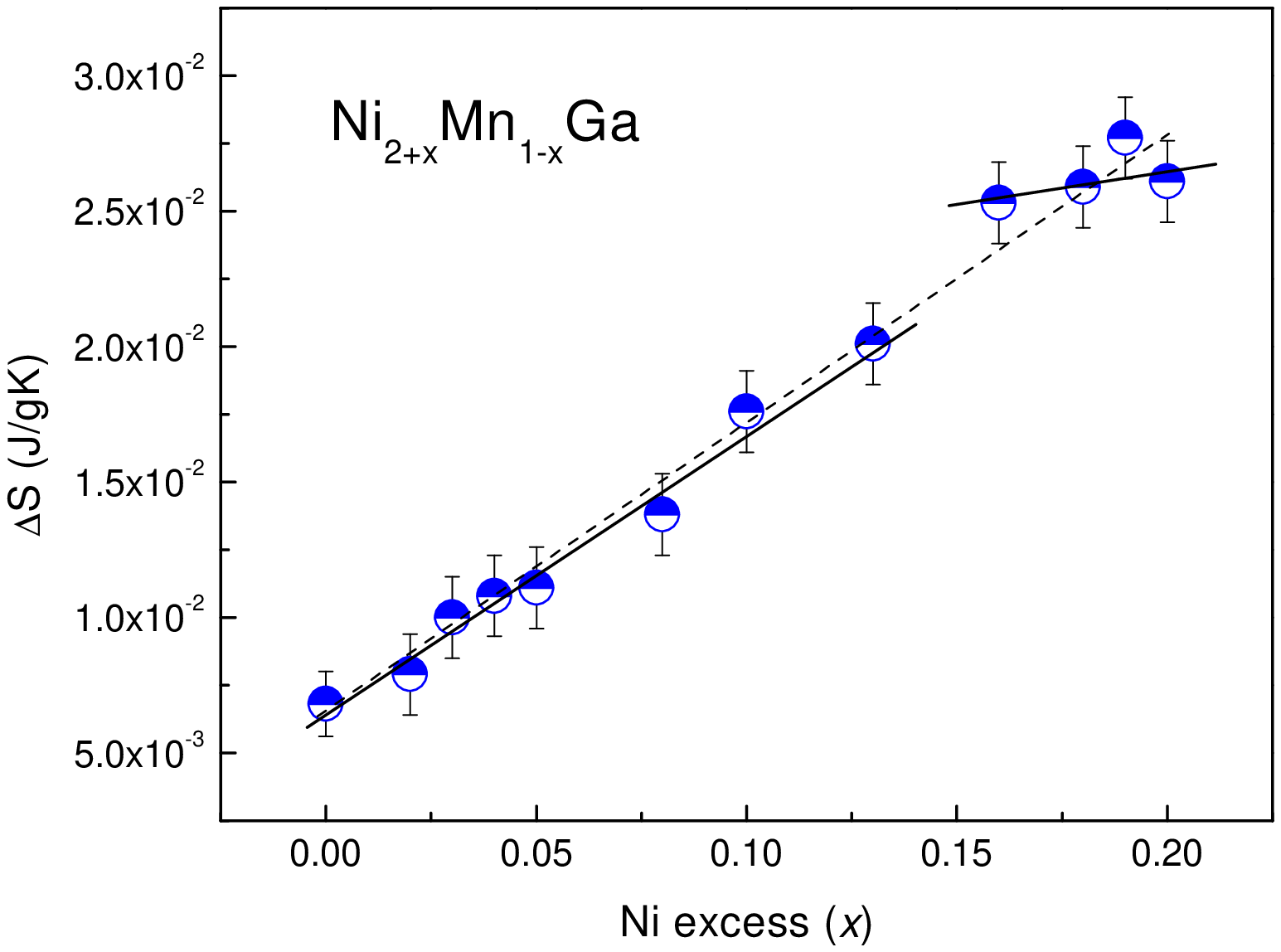}
\caption{The entropy change at the martensitic transformation in
Ni$_{2+x}$Mn$_{1-x}$Ga alloys as a function of Ni excess $x$. The
solid lines are linear fits to the data.}
\end{center}
\end{figure}

Since configuration contributions to the entropy change are absent
in the case of martensitic transformations, it is customary to
consider that $\Delta S$ has three main contributions:

$$
   \Delta S = \Delta S_{vib} + \Delta S_{el} + \Delta S_{mag},
   $$

\noindent where $\Delta S_{vib}$ is the vibrational contribution,
$\Delta S_{el}$ is the contribution of the conduction electrons,
and $\Delta S_{mag}$ is the contribution from magnetic subsystem.
Although specific heat measurements of Ni-Mn-Ga alloys at a low
temperature have not been performed, it can be expected,
nevertheless, that the electronic contribution $\Delta S_{el}$ to
the entropy change in Ni$_{2+x}$Mn$_{1-x}$Ga alloys is small. This
assumption is supported by the measurements of specific heat at
low temperatures for several ferromagnetic X$_2$MnSn (X = Co, Ni,
Pd, and Cu) Heusler alloys.~\cite{8-fraga} Thus, the increase in
$\Delta S$ with the deviation from the stoichiometry in
Ni$_{2+x}$Mn$_{1-x}$Ga is likely due to the $\Delta S_{vib}$ and
$\Delta S_{mag}$ terms.

\begin{table}
\caption{Heat exchanged upon reverse ($Q^{L\to H}$) and direct
($Q^{H\to L}$)martensitic transformation, enthalpy ($\Delta H$)
and entropy ($\Delta S$) changes for Ni$_{2+x}$Mn$_{1-x}$Ga
alloys.}
\begin{ruledtabular}
\begin{tabular}{ccccc}

Alloy & $Q^{L\to H}$ (J/g) & $Q^{H\to L}$ (J/g) & $\Delta H$ (J/g)
& $\Delta S$ (mJ/gK) \\

\hline

\hline

$x = 0$ & 1.39 & - 1.34 & 1.365 &  6.8 \\

\hline

$x$ = 0.02 & 1.7 & - 1.87 & 1.785 &  7.9 \\

\hline

$x$ = 0.03 & 2.38 &   - 2.3  & 2.34  &  10 \\

\hline

$x$ = 0.04 & 2.6 & - 2.57 & 2.585 &  10.8 \\

\hline

$x$ = 0.05 & 2.72  &  - 2.72 & 2.72  &  11.1 \\

\hline

$x$ = 0.08 & 3.65  &  - 3.8 &  3.725  & 13.8 \\

\hline

$x$ = 0.10 & 4.69  &  - 5.09 & 4.89  &  17.6 \\

\hline

$x$ = 0.13 & 5.71 &   - 5.57 & 5.64  &  20.1 \\

\hline

$x$ = 0.16 & 7.96  &  - 7.74 & 7.85  &  25.3 \\

\hline

$x$ = 0.18 & 8.65  &  - 8.57 & 8.61  &  25.9 \\

\hline

$x$ = 0.19 & 9.41  &  - 9.59 & 9.5 & 27.7 \\

\hline

$x$ = 0.20 & 9.08  &  - 8.86 & 8.97 &   26.1 \\

\end{tabular}
\end{ruledtabular}
\end{table}

An analysis of  $\Delta S_{vib}$ contribution to the entropy
change of Cu-based shape memory alloys showed that the vibration
contribution depends on the elastic anisotropy at the
transformation temperature.~\cite{1-planes} The authors found that
for a given crystal structure of the martensitic phase, the
elastic anisotropy constant $A$ at $M_s$ does not depend on
composition, which means that vibration contribution to $\Delta S$
remains constant for all of the composition studied. In the case
of Ni$_{2+x}$Mn$_{1-x}$Ga alloys, data on elastic anisotropy are
absent. The observed increase of $\Delta S$ in
Ni$_{2+x}$Mn$_{1-x}$Ga can indicate that elastic anisotropy
depends on composition. However, since Debye temperature did not
change significantly with composition,~\cite{9-matsumoto} it is
more likely, that composition dependence of $\Delta S$ is
accounted for by the magnetic contribution $\Delta S_{mag}$.

The fact that entropy change in Ni-Mn-Ga alloys depends on
composition has already been mentioned in Ref.~10, where Ni-Mn-Ga
alloys were divided into three groups according to their
transformation behavior. The authors found that alloys with a low
martensitic transformation temperature $M_s$ have low values of
$\Delta S$, whereas alloys with a high $M_s$ are characterized by
higher values of the entropy change. Their observations agree with
the results of our study.

Since the crystal structure of the martensitic phase of Ni-Mn-Ga
depends on composition,~\cite{6-pons} an analysis of $\Delta S$ as
a function of Ni excess $x$ is worth performing. From the data
shown in Fig.~2, it is difficult to draw an unambiguous conclusion
because the composition dependence of $\Delta S$ can be
approximated in the whole studied interval of $x$ as shown by the
dashed line. However, we suggest that the entropy change has
different composition dependencies in concentration intervals $0
\le x \le 0.13$ and $0.16 \le x \le 0.20$ as shown in Fig.~2 by
the solid lines. The alloys with $x \ge 0.16$ ($M_s > 300$~K) are
expected to have seven-layered martensitic structure, as evident
from their high martensitic transformation temperatures and
unusual behavior of resistivity,~\cite{5-vas,11-kvv} whereas the
$0 \le x < 0.16$ alloys undergo structural transformation to the
five-layered martensitic structure. Since different martensitic
phases have different densities of vibrational states, this should
lead to discontinuity of the composition dependence of $\Delta S$
as the martensite of the Ni$_{2+x}$Mn$_{1-x}$Ga alloys changes its
modulation. If this is the case, the alloys with seven-layered
modulation of the martensitic phase are characterized by a higher
$\Delta S$ as compared to the alloys with five-layered modulation
(Fig.~2).

\bigskip


In this article we have studied the entropy change at the
martensitic transformation in Ni$_{2+x}$Mn$_{1-x}$Ga ($x = 0 -
0.20$) alloys. The lowest value of the entropy change, $\Delta S =
6.8$~mJ/gK, was found for the stoichiometric Ni$_2$MnGa. Upon
substitution of Mn for Ni, $\Delta S$ significantly increases up
to $\sim 26$~mJ/gK in alloys with $x > 0.16$. The increase in
$\Delta S$ is presumably due to the magnetic contribution. It is
suggested that the change in modulation of the martensitic phase
results in discontinuity of the composition dependence of $\Delta
S$. This assumption, however, requires further systematic studies
of thermodynamic properties of Ni-Mn-Ga alloys.

\section*{Acknowledgments}

This work was partially supported by a Grant-in-Aid from Izumi
Science and Technology Foundation. One of the authors (V.V.K.)
gratefully acknowledges the Japan Society for the Promotion of
Science (JSPS) for a Fellowship Award.

\end{document}